\documentclass[journal]{IEEEtran}

\usepackage{cite}
\usepackage{graphicx}
\usepackage[table]{xcolor}
\usepackage{csquotes}
\usepackage{array}
\usepackage{multirow}
\usepackage{arydshln}
\usepackage{pdfpages}
\usepackage{xurl}
\usepackage[hidelinks]{hyperref}

\newcommand{\bibdoi}[1]{doi: \href{https://doi.org/#1}{\nolinkurl{#1}}}

\begin{document}

\title{Complying with Cybersecurity Legislation: NIS2 Challenges and Industry Case Study}

\author{
	\IEEEauthorblockN{
		Ricardo Galvão\IEEEauthorrefmark{1},
		Miguel Coelho\IEEEauthorrefmark{2},
		João R. Campos\IEEEauthorrefmark{1}
	}
	\IEEEauthorblockA{\\\IEEEauthorrefmark{1}
		Department of Informatics Engineering, CISUC/LASI\\
		University of Coimbra, Coimbra, Portugal\\
		\{ricardogalvao@student.dei.uc.pt, jrcampos@dei.uc.pt\}
	}
	\IEEEauthorblockA{\\\IEEEauthorrefmark{2}
		Altri, Porto, Portugal\\
		miguel.coelho@altri.pt
	}
	\thanks{Corresponding author: João R. Campos (jrcampos@dei.uc.pt).}
	\thanks{This work was funded by national funds through FCT -- Foundation for Science and Technology, I.P., under research unit UID/00326 -- Centre for Informatics and Systems of the University of Coimbra, and Agenda ``NEXUS -- Pacto de Inovação Transição Verde e Digital para Transportes, Logística e Mobilidade'', financed by the Portuguese Recovery and Resilience Plan (PRR), project no. C645112083-00000059 (investment project no.~53).}
}

\maketitle

\begin{abstract}
Given the ever-increasing dependency on interconnected digital systems, cybersecurity has become a major concern. To protect markets and societies, governmental entities are developing legislation to enhance cyber resilience across sectors, introducing new technical and legal requirements. The \textit{NIS2 Directive}, a revised version of the European-wide cybersecurity legislation, entered into force on 16 January 2023. Member States were required to adopt and publish transposing measures by 17 October 2024 and apply them from 18 October 2024. Specific obligations for essential and important entities are set out in national legislation. At the same time, other organizations, while not legally bound, seek to improve their cybersecurity maturity in alignment with EU-wide advances, for which NIS2 serves as a key reference. This paper presents the work conducted within a large European enterprise to align itself with the \textit{NIS2 Directive}, focusing on how organizations can analyze and interpret its provisions to drive concrete technical and organizational improvements. A systematic methodology is applied to frame the company within the NIS2 context, identify relevant requirements, and derive actionable measures. Beyond outlining the technical steps taken, the work highlights the challenges of translating legislative requirements into practical implementations, emphasizing the use of existing IT resources while accounting for key legal aspects. The paper concludes with a discussion and reusable artifacts addressing critical areas such as \textit{Supply Chains and Third Parties} and \textit{Zero-Trust Principles}, illustrating how legislative instruments such as NIS2 can be translated into practical mechanisms to strengthen cybersecurity strategies.
\end{abstract}


\section{Introduction}
Digital services are increasingly interconnected, enabling a level of data correlation and automation that was not possible before. The perception of information and control systems as valuable assets has become inevitable, and cyberspace has turned into a highly appealing target for malicious actors. Studies report a considerable increase in cybercrime, expected to cost the world \$23 trillion in 2027, an increase of 175\% from 2022 \cite{cybercrimestats}. 2023 saw a 72\% increase in the number of breaches compared with 2021~\cite{forbes}. For example, the MOVEit 2024 breach affected 77 million individuals and more than 2,600 companies, causing over \$12 billion in damage \cite{bluefin}. Similarly, the CDK 2024 breach affected thousands of dealerships, reportedly costing tens of millions of dollars \cite{crn}. The impact of these incidents extends beyond business disruption and financial loss, it can directly affect citizens and institutions connected to the targeted entities, or indirectly compromise the availability of critical services such as healthcare and communications. 

IBM reported a global average data breach cost of USD~\$4.88 million in 2024, up 10\% from 2023~\cite{ibmBreach2024}. The European Union Agency for Cybersecurity (ENISA) identified vulnerability exploitation as a common way attackers gained access and reported 42,595 new vulnerability disclosures from July 2024 through June 2025, a 27\% increase over the preceding year~\cite{enisaThreatLandscape2}.

To counter the growing threat to the economy and society, significant effort has been devoted to developing legislation aimed at improving cybersecurity and resilience \cite{legislation_2024}. The European Union (EU) has taken a leading role, establishing a comprehensive framework of initiatives such as the Cyber Resilience Act (CRA)~\cite{cra}, which introduces rules to ensure safer hardware and software, and the \textit{NIS2 Directive}\cite{nis2_legislation}, a revised and strengthened version of the previous NIS Directive \cite{nis_legislation}. As a directive, NIS2 sets out strategic goals and minimum cybersecurity measures that Member States must transpose into national law. National transposing laws specify country-level legal obligations. This study focuses on the \textit{NIS2 Directive} itself as the overarching European framework. The analysis therefore targets how its principles can be interpreted and operationalized from a technical and organizational standpoint, rather than examining national legal implementations. Importantly, its impact extends beyond entities legally bound by it, many organizations aim to align with NIS2 to improve their cybersecurity maturity and resilience.

New legislation inevitably drives significant changes, yet translating legal provisions into effective technical and organizational measures remains challenging. The legal text of directives is typically written in terms that are difficult to interpret from a management or technical standpoint, and does not directly map to concrete architectures or security controls. Furthermore, numerous organizations such as the National Institute of Standards and Technology (NIST), the Open Worldwide Application Security Project (OWASP), and ENISA publish complementary frameworks and guidance, including the NIST Cybersecurity Framework~\cite{nistFramework}, the OWASP Cheat Sheet Series~\cite{owaspCheatSeries}, and ENISA guidance on security measures~\cite{enisaTechGuideSecMeasures}, making it difficult for organizations to determine which resources to prioritize. Despite the growing availability of cybersecurity information, adoption is still hindered by limited resources and expertise \cite{cybercrimestats}.

In this context, analyzing the directive itself provides an opportunity to establish a proactive, structured mechanism for interpreting high-level legislative principles and transforming them into actionable security practices before specific national requirements are defined.

In this paper, we present work conducted within a large European manufacturing enterprise to analyze and operationalize the \textit{NIS2 Directive} from a technical perspective. Our approach focuses on how organizations can systematically interpret and apply the directive’s provisions to guide technical and organizational improvements. First, a comprehensive study was performed to contextualize the company within the NIS2, identifying relevant requirements while considering its existing structure, processes, and cybersecurity maturity. Based on a well-defined analysis and prioritization process, the most critical topics were explored in depth to produce supporting documents and assets that guide the company’s cybersecurity strategy in alignment with NIS2 objectives.

This paper makes two key contributions. First, it presents a \textbf{methodology to analyze how an organization can operationalize NIS2 principles to inform technical and organizational decision-making and demonstrates its application in a real-world context, highlighting the challenges of translating legal obligations into actionable security measures}. Second, it delivers \textbf{a set of publicly available artifacts} covering essential areas such as \textit{Secure Software Development Life Cycle (SDLC)}, \textit{Supply Chains and Third Parties}, and \textit{Zero-Trust Principles}. These outputs provide actionable guidance for organizations aiming to enhance cybersecurity maturity and align with EU-wide resilience goals.

The remainder of the paper is organized as follows. \textit{Section \ref{sec:background}} reviews relevant cybersecurity legislation and related work. \textit{Section \ref{sec:methodology}} presents the proposed methodology, while \textit{Sections \ref{sec:case_study_ass}} and \textit{\ref{sec:case_study_dev}} describe its application and results. \textit{Section \ref{sec:discussion}} discusses the main findings, and \textit{Section \ref{sec:conclusion}} concludes the paper and suggests directions for future work.

\section{Background and Related Work}
\label{sec:background}

In this section we overview existing cybersecurity legislation and discuss similar related works.

\subsection{Cybersecurity Legislation}

Within the EU, cybersecurity is managed by the European Union Agency for Cybersecurity (ENISA). The EU Cybersecurity Strategy \cite{2.1_10}, revised in 2020 \cite{2.1_11}, focuses on the development of a global and open cyberspace through cooperation, ensuring operational capacity to prevent, deter and responding to the incidents, and achieve resilience while maintaining technological sovereignty and leadership at regional and international level. 

The CRA \cite{2.1_13} was approved in 2024 as a regulation on cybersecurity requirements for products with digital elements. It focuses on the widespread vulnerabilities combined with an insufficient provision of security updates to fix them, compounded by the lack of user knowledge and available information to allow informed decisions when choosing products with adequate security properties. 

The NIS2 Directive entered into force on 16 January 2023 (Article~45). Under Article~41, Member States were required to adopt and publish their transposing measures by 17 October 2024 and apply them from 18 October 2024~\cite{nis2_legislation}. It strengthens cybersecurity requirements for critical infrastructure and essential services, expanding its scope over NIS \cite{nis_legislation} to include more sectors and organizations. It defines stricter security measures, incident reporting, and risk management practices while imposing tougher penalties for non-compliance. It focuses on novel and actual threats such as supply chains and third parties which have become some of the main threats and attack vectors used in advanced campaigns.

In the field of data protection, the General Data Protection Regulation (GDPR) \cite{2.1_20} defines obligations to all organizations that target or collect data related to EU citizens. The Data Governance Act (DGA) \cite{DGA} provides a framework to regulate the re-use of publicly/held and protected data while enhancing trust in voluntary data sharing. The eIDAS Regulation \cite{eIDAS} ensures that both people and business can use their own national electronic identification schemes to access public services available in other EU countries and the European Data Act \cite{EUdataAct} aims to boost the usage of data in the advent of the Internet of Things (IoT) and its potential for an exponential growth of data volume. Similar to the US sector-specific legislation also exists, such as Digital Operational Resilience Act (DORA) \cite{dora} and Digital Services Act (DSA) \cite{dsa}.

\subsection{Frameworks}

Given the growing relevance of cybersecurity various standards, processes, and guidelines have been proposed over the years by various prominent entities such as NIST, OWASP, ENISA, CIS, and ISO/IEC. These resources are often made available targeting specific domains, such as IoT and AI.

NIST Cybersecurity Framework 2.0 is a general cybersecurity framework  \cite{nistFramework}. It represents an updated version over version 1.1 with enhanced guidance and recommendations for managing and improving cybersecurity risk. It reflects the evolving landscape of cybersecurity threats and incorporates feedback from stakeholders to strengthen organizations' cybersecurity practices. NIST also proposed NIST SP 800-95: Guide to Secure Web Services \cite{nistSP800-95} which provides guidelines on secure web application development. It addresses security considerations for web services and offers recommendations for developers and administrators to enhance the security of web applications and web services. Other NIST standards and resources exist for other domains, such as mobile \cite{nistSP800-163} and IoT \cite{nistIR8320}. 

ISO 27002:2022 \cite{iso27002:2022} is an international standard that provides a set of guidelines and best practices for information security management. It offers a comprehensive framework for organizations to establish, implement, maintain, and continually improve their information security management systems, closely aligned to ISO 27001:2022. Other standards also exist (e.g., ISO/IEC 15408-1:2022 \cite{iso15408:2022}) some specific to different application domains (e.g., IoT~\cite{iso27400:2022}). 

CIS Critical Security Controls v8 \cite{cisControls}
are a set of prioritized cybersecurity best practices and defensive measures designed to help organizations mitigate the most common and damaging cyber threats. The controls cover various aspects of cybersecurity, including asset management, vulnerability assessment, access control, malware defense, and incident response, over multiple domains such as cloud and mobile technologies. By implementing the CIS Critical Security Controls, organizations can better protect their systems, networks, and data from a wide range of cyber attacks. CIS Community Defense Model also exists \cite{cisDefense}, a framework aimed at enhancing cybersecurity through collaboration and information sharing among various stakeholders.

OWASP is an organization that works to improve the security of software and provides an extensive set of resources. Some of the most well-known are the OWASP Cheat Sheet Series (OCSS) \cite{owaspCheatSeries} 
is a collection of practical security-related resources for software developers and security professionals. It includes cheat sheets and concise guides covering various aspects of application security, offering quick references for secure coding practices. Another relevant asset provided by OWASP are their Top 10 rankings. The OWASP Top 10 \cite{owasptop10} addresses the most critical security risks to web applications, while the OWASP API Security Top 10 \cite{owaspAPIsecurityTop10} aims to raise awareness about the most critical security risks associated with API usage. Other Top 10 rankings also exist for various domains such IoT \cite{owaspIOTtop10}
and Machine Learning \cite{owaspMachinelearningTop10}. OWASP also provides OWASP Application Security Verification Standard (ASVS) \cite{owaspASVS}, a framework of security requirements for designing, developing, and testing modern web applications and web services. The OWASP Web Security Testing Guide \cite{owaspWEBsecurTest} is an in-depth guide to evaluating the security of web applications and web services. Similar resources also exist for mobile applications \cite{owaspMASVS,owaspMASTG} and IoT \cite{owaspIOTtestGuide,owaspIOTisvs}.

ENISA Technical Guideline on Security Measures \cite{enisaTechGuide} outlines specific technical guidelines and recommendations for implementing security measures. This document covers various aspects of cybersecurity, providing detailed guidance for organizations to enhance their security posture. Various other resources from ENISA also exist, such as National Cybersecurity Strategies Guidelines \& tools \cite{enisaTechGuideSecMeasures} and Secure Software Engineering Initiatives \cite{enisaSecureSoftwareInitiatives}.

Various other resources have also been proposed such as	Cybersecurity Capability Maturity Model (C2M2) \cite{C2M2} and Cybersecurity Maturity Model Certification (CMMC) 2.0 \cite{CMMC}.

One of the main challenges, which is also highlighted in recent legislation \cite{cra}, is that there are currently too many frameworks with different perspectives and implementations, but with significant overlaps among them. Moreover they are often extensive documents, which makes both the process of defining which solutions to use and how to put them to practice a challenging task.

\subsection{Related Works}

With the entry into force of the NIS2 Directive and its transposition into national laws by EU Member States, a growing body of literature has emerged analyzing its components, implications, and implementation challenges. These studies collectively emphasize the evolving cybersecurity governance landscape in Europe and the difficulties organizations face in adapting to the new requirements.

Early works, such as \cite{2.3_65}, provide an overview of the process of adopting new regulatory frameworks, emphasizing the need to understand the framework in depth and prepare organizations for change. Lindström et al.~\cite{2.3_67} highlight the importance of maintaining well-documented cybersecurity policies, particularly in operational contexts.
Further analyses by Maglaras et al. \cite{2.3_68}, Shukla et al. \cite{2.3_69}, Wallis and Johnson~\cite{2.3_70}, Szczepaniuk et al. \cite{2.3_71}, and Ubowska and Królikowski~\cite{2.3_72} examine the implementation of the first NIS Directive, identifying challenges such as the complexity of overlapping legislation, organizational resistance to change, and communication barriers with decision-makers. These works also stress the necessity for holistic and adaptable approaches to cybersecurity management.

From a complementary perspective, Rawindaran et al. \cite{2.3_73} and Mirtsch et al. \cite{2.3_75} explore the practical barriers faced by organizations in translating legal obligations into actionable cybersecurity measures. Their findings underscore the persistent gap between high-level regulatory intent and its low-level operational implementation.

More recent studies focus directly on NIS2 and its expected impact. Singh \cite{11_10.1} provides a comparative review of the directive, assessing its implications for affected organizations and contrasting them with the UK’s evolving cybersecurity framework post-Brexit. Van 't Schip~\cite{11_10.2} addresses the expanded supply chain considerations introduced by NIS2, noting that while the directive strengthens supply chain accountability, it may overlook the technical complexity of cybersecurity ecosystems, particularly concerning the Internet of Things.

Avramidou et al.~\cite{11_10.3} raise concerns regarding overlapping responsibilities and the lack of clarity in the classification of critical entities under current regulatory interpretations. Busetti and Scanni~\cite{11_10.4} find that incident reporting helps prompt a rapid response and may inform policy improvements, although limited capacity and reluctance to report can hinder that learning. Schmitz-Berndt~\cite{11_10.5} identifies NIS2 as a step toward more efficient and transparent reporting.

Across these contributions, a recurring challenge emerges: the translation of legislative requirements into concrete organizational practices without compromising business continuity. While NIS2 represents a significant regulatory advancement, its practical implementation remains underexplored. This study seeks to address that gap by providing a structured set of resources to support organizations in defining and operationalizing cybersecurity strategies aligned with NIS2.
\section{Operationalizing the \textit{NIS2 Directive}}
\label{sec:methodology}

The \textit{NIS2 Directive} introduces a strengthened European framework for cybersecurity governance and resilience. While its legal obligations apply to specific essential and important entities, its principles offer valuable guidance for any organization seeking to enhance its cybersecurity maturity. Viewing NIS2 as a strategic framework rather than a compliance checklist allows organizations to anticipate future expectations and align with recognized best practices.

\begin{figure*}[t]
	\centering
	\includegraphics[width=1\linewidth]{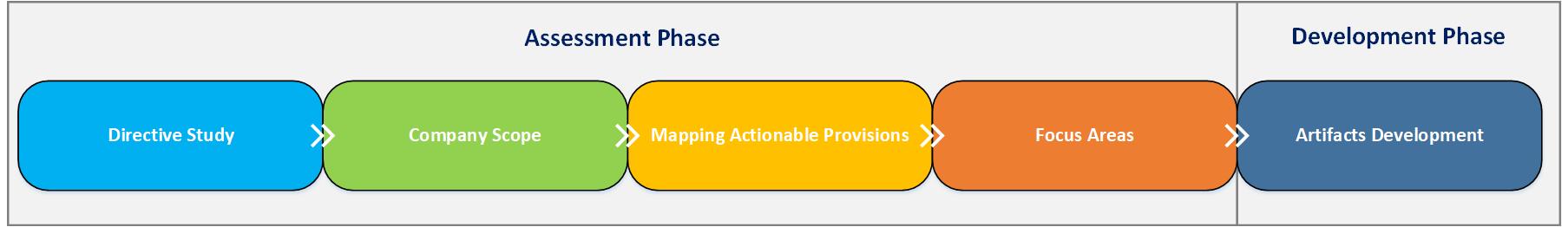}
	\caption{Framework for Operationalizing the \textit{NIS2 Directive}} 
	\label{fig:figRoadMap2.0}
\end{figure*}

This section presents the methodology used to interpret and operationalize the \textit{NIS2 Directive} within a large European private enterprise, identifying how its provisions can be translated into concrete technical and organizational measures. The methodology, adapted from Beasley's guidance on implementing the NIST Cybersecurity Framework~\cite{2.3_65}, is depicted in \textit{Figure \ref{fig:figRoadMap2.0}} and is composed of five sequential stages grouped into two phases: an \textit{Assessment Phase}, where the directive and the organizational context are analyzed, and a \textit{Development Phase}, where targeted cybersecurity actions and artifacts are produced:

\begin{enumerate}
    \item \textbf{Directive Study} – Systematic analysis and indexing of the directive to identify relevant provisions, definitions, and cybersecurity themes
    \item \textbf{Company Scope} – Characterization of the organization to determine how it fits within the NIS2 context
    \item \textbf{NIS2 Provisions Mapping} – Identification and categorization of NIS2 provisions with potential technical and organizational implications
    \item \textbf{Focus Areas} – Selection of key cybersecurity domains requiring further development or reinforcement
    \item \textbf{Artifacts Development} – Creation of practical resources to support the organization’s cybersecurity strategy and alignment with NIS2 objectives
\end{enumerate}

Given the expanded scope of NIS2, its principles are expected to impact a wide range of organizations, regardless of whether they are formally designated as essential or important entities. Accordingly, the proposed methodology focuses on the systematic analysis and conversion of NIS2 provisions into actionable cybersecurity measures, enabling organizations to strengthen their posture both in anticipation of and in alignment with the directive’s objectives. This work does not address legal interpretation or compliance verification; rather, it adopts a technical and organizational perspective, using NIS2 as a high-level framework for deriving practical cybersecurity enhancements.

\subsection{Directive Study}
The first step involves a systematic examination of the \textit{NIS2 Directive} to identify its structure, terminology, and key thematic areas relevant to technical and organizational implementation. This mapping facilitates communication across management and IT teams, ensuring a shared understanding of the directive’s scope and intent.

\subsection{Company Scope}
The Company Scope stage uses the organization's activities, size, and other relevant characteristics to assess which NIS2 provisions may apply. The assessment draws on the scope and classification criteria in Articles~2 and~3 and on the sectoral descriptions in Annexes~I and~II~\cite{nis2_legislation}. Where the assessment raises questions about applicability or classification, legal input can help identify the provisions to consider in the subsequent technical mapping.

\subsection{Mapping Actionable Provisions}
Using the scope assessment from the previous stage, the next step involves identifying and categorizing the provisions of the \textit{NIS2 Directive} that can be translated into actionable cybersecurity and organizational measures. This requires a systematic analysis of the directive’s articles and recitals to extract relevant principles, requirements, and technical expectations, resulting in a structured mapping of its contents. The analysis also extends to referenced external documents to identify additional standards and guidelines that complement the directive and inform its practical implementation.

One of the main challenges in this stage is subjectivity, since some of the considerations in NIS2, although not specifically targeting entities, advise Member States to promote the development of certain practices/processes. Such suggestions might be transposed to obligations and thus should also be considered.
The main outcome of this stage is a detailed catalog with all the relevant cybersecurity items and topics, that can be perceived as relevant technical and organizational measures to be addressed by the company. 

\subsection{Focus Areas}
The resulting catalog of the previous stage should then be checked against the company's business process and its cybersecurity infrastructure. The IT team of the entity under study should be included in the process, through the use of semi-structured interviews and/or reviews of materials and documentation. It is used to cross-analyze against the entity's current practices and cybersecurity maturity level. This aims to identify key areas and topics where further research and development is necessary for aligning with NIS2. Depending on the level of detail/information and type of interviews, ethical considerations might be necessary to protect the participants and obtain the necessary permissions.

The assessment is carried out collaboratively with the organization’s IT and management teams to ensure that identified focus areas reflect both operational priorities and the intent of the \textit{NIS2 Directive}. The assessment considers several factors, including the organization’s available resources, the status and maturity of ongoing cybersecurity initiatives, and their alignment with the principles and objectives of the \textit{NIS2 Directive}. Levels of training and cybersecurity awareness among staff are also analyzed, as they directly influence decision-making efficiency and resource management. Limited understanding of the directive’s implications or of possible technical responses is identified as both a risk and a challenge at this stage. To support prioritization, additional data from reliable sources—such as threat-landscape reports developed by competent authorities—are integrated to provide evidence-based guidance on which areas to address first.

The expected result is a list of topics classified by priority. This prioritization will assist the organization in managing its resources while maximizing its return. The priorities are established through a set of metrics,  defined considering the cybersecurity context of the company and its goals. Priority is given to measures with high potential to become legal requirements after NIS2's transposition, followed by the guidelines and more generic recommendations. Areas not recently developed in the company or previously marked by it as potential improvements will take preference over topics already being covered or targeted by ongoing projects. 

\subsection{Artifacts Development}
The goal of the previous stages is to identify the most relevant cybersecurity domains requiring development or reinforcement to align with the \textit{NIS2 Directive}. To support this process, a set of artifacts was developed to translate the directive’s intent into practical guidance and resources for technical teams. These artifacts will support, determine, and provide a path for the changes that need to be made. 

Research and analysis of both state-of-the-art and state-of-practice needs to be conducted. Documentation is to be developed to form a library that will guide the company's efforts in the different identified relevant areas. One challenge that arises at this stage is that the domains of cybersecurity are wide with potential for interconnections and complexity. A single technology or ICT product can have many different roles, architectures or deployed configurations, depending on the security aspects being considered (confidentiality, availability or integrity). A single concept may be composed of a variety of technologies, architectures, frameworks and other related terms that can also vary individually depending on the context being considered. 

To guide the research process, it is important to keep a close relationship with the stakeholders involved in the company. At the beginning of the development of each defined topic, a checkpoint session (e.g., in the form of semi-structured interviews) should be conducted, aiming to assess and review in detail the company's awareness and expectations concerning the topic's domain research. This information allows tailoring the outcome of the research, either focusing on more technical/concrete solutions (e.g., a product, a framework, a standard) or a broader overview of the topic's domain. In case of topics already in development within the company it is possible to focus on a more specific/technical approach. On the other hand, broader topics are addressed through a more general perspective to ensure full awareness of all the potentially relevant aspects, with the possibility for further specification as its cybersecurity maturity increases.  

To provide consistency, the artifacts produced should present a similar overall structure. Although the needs of the artifacts can vary between topics and the structure can be tailored for each organization, a three-part structure can serve as a starting point. The artifact starts by providing an introduction and considerations related to the topic, for the purpose of introducing the reader to the domain being discussed. It aims to ensure that the terms and notions used are clear, providing awareness about the relevance of it and, when applicable, the potential extension of the topic's domain. The second element is the identification of existing standards and frameworks related to the topic. The third element is the presentation of available solutions and/or sources of relevant material related to the topic. To ensure relevance in the scope of NIS2 as a European directive priority was overall given to EU sources from national and international entities such as CNCS and ENISA.

By combining insights from NIS2 with existing standards and best practices, these artifacts enable organizations to operationalize legislative principles in their cybersecurity strategies, ensuring consistency with EU-wide resilience objectives.

\section{Case Study: Assessment Phase}
\label{sec:case_study_ass}

The case study focuses on Altri, a leading European enterprise in the production of cellulosic fibers and forest management. Established in February 2005, among others, the group is responsible for forest management and three cellulosic fibers mill units. The Altri infrastructure, as a group, is comprised of several factories/mills. The analysis focuses on its main factory unit, Celbi, selected for its technological profile and strategic relevance.

This work resulted from a collaboration initiated by Celbi's IT department and our University's cybersecurity group. Celbi, with its high revenue and valuable assets, aims to proactively enhance its cybersecurity posture by using legislation as a strategic driver for improvement while preparing for upcoming regulatory requirements. The collaboration gave the research team access to Celbi's internal processes for the assessment.

To ensure ethical and security compliance, all insights and results were reviewed collaboratively with Celbi’s IT department and relevant stakeholders. Maintaining a sustainable communication flow without interfering with the company’s operations was essential. This was achieved through semi-structured one-hour group interviews with three main objectives: presenting insights, outlining next steps, and addressing questions or requirements. These sessions ensured that the study remained aligned with Celbi’s current status, needs, and objectives. No identifiable data were collected during the interviews, which prioritized consistent communication, progress awareness, and actionable outcomes.

To provide a further detailed description of the documents produced we decided to separate the two phases of our methodology in different sections. The outputs of the \enquote{Development Phase} will be presented in \textit{Section \ref{sec:case_study_dev}}. This section focuses on the \enquote{Assessment Phase}, which comprises stages 1 to 4, and focuses on collecting answers to key questions that will help properly measure and plan the necessary work to prepare the company for NIS2 while focusing on its respective needs. 

\subsection{Directive Study}
\label{sec:directiveStdy}
The \textit{NIS2 Directive} document starts by enumerating key points and considerations along 144 recitals. These recitals aim not only to specify the context and objectives of the directive but also to clarify many important considerations such as allocated tasks and responsibilities, approaches in case of conflict, internal procedures for assistance and reviews, among others. Following this introduction there are nine chapters and three auxiliary appendixes.

The case study used the EU directive as its reference while Portuguese transposition was pending. Portugal subsequently transposed NIS2 through Decree-Law No.~125/2025, published on 4 December 2025 and in force from 3 April 2026~\cite{portugal_nis2_2025}. The assessment reflects the provisions available at the time of the study.

The outcome of this stage is a set of tables summarising the 144 initial recitals combined with an index of the chapters' contents to facilitate the analysis of the document and interactions with relevant stakeholders.

\subsection{Company Scope}
\label{sec:company}

Understanding how an organization fits within the general scope of the \textit{NIS2 Directive} is a necessary first step to contextualize its cybersecurity priorities and the relevance of its provisions. In this study, a high-level assessment was conducted to identify which NIS2 principles and domains are most applicable to Celbi’s operational and technological context. Rather than performing an exhaustive legal classification, the focus was on interpreting the directive’s structure and terminology to anticipate which requirements could influence the company’s cybersecurity strategy. This assessment was reviewed in collaboration with Celbi’s legal department to ensure that the interpretation remained consistent with legal definitions.

For planning purposes, Celbi chose to assess the more stringent scenario in which it would be treated as an \textit{essential entity}. The study used that assumption to select provisions for technical review. Celbi's formal classification under Portuguese law was outside the scope of the assessment.

\subsection{Mapping Actionable Provisions}
\label{sec:NIS2topics}

Using Celbi's planning assumption, the authors assessed the NIS2 security requirements and recommendations relevant to that scenario. As seen in the directive's study (\textit{Section \ref{sec:directiveStdy}}), NIS2's contents are not exclusively directed to entities, containing other subjects not relevant for this step such guidelines for different workgroups, tasks for governmental bodies, procedures between EU relevant entities (to name a few). To gather only the relevant data for this stage, the assessment was divided in two steps. 

For the first step, a focused search using a specific set of keywords (e.g., \enquote{entity}, \enquote{entities}, was conducted against the directive to locate all the recitals that fit the Celbi classification and operational context. The matching recitals were then analyzed to exclude those not targeting entities such Celbi. The second step consisted of a full reading of the directive and cross-validation with the previously identified recitals to ensure that no measures or considerations were missed due to terms that were not included in the first stage.

Given the diversified nature of the contents present in the recitals (e.g., definition of a specific measure or recommendation, clarifying a term, or presenting relevant considerations on a topic), they were organized into groups. Moreover, although some recitals cover in detail a single topic, others are more generic. For those cases, it was used the main drive or aim of the different topics as criteria to attribute a group (for example, a paragraph covering notes regarding multiple systems without recommending an action would fall into the group "considerations"). These groups act as an organizational framework for mapping NIS2 recommendations, adaptable to business context, improving the organization and structure of collected content into a detailed map, which will be the input for the next stage. Inspired by the NIST framework, it breaks down complex security domains into manageable components, enabling focused analysis while maintaining a global view. It is worth noting that these groups are products of this study, not reflecting any official categorization defined by the directive: 

\begin{itemize}
	\item \textbf{Considerations} – Covers aspects that, although not specifying any measure or security obligation towards the entities, still contain relevant information such as clarifications on a specific term/definition, or detailing informational aspects such, for example, how responsibilities of relevant bodies are distributed on a specific flow of actions.
	\item \textbf{Scope} – recitals that specify the scope of the measures or the respective components that should be considered or be covered by them;   
	\item \textbf{Measures} – recitals that specify measures required to apply, or guidelines that may result in such measures; 
	\item \textbf{Supply Chains} – Measures and considerations related to entities direct suppliers and service providers;
	\item \textbf{Third Parties} – Details focusing on the relations that entities establish with third parties in broader scopes (for example, relations with other institutes for the purpose of research cooperation);
	\item \textbf{Reporting} – recitals related to reporting obligations such as dates, procedures that entities need to follow, contents that should be included, etc;
\end{itemize}

\textit{Table \ref{fig:relevantParagraphSummary}} presents a summary of the collected measures used to build a full detailed map that was then delivered to the company's cybersecurity team.

\begin{table}[ht]
	\renewcommand{\arraystretch}{1.5}
	\centering
	\caption{Summary of NIS2 measures and considerations relevant to Celbi. Standalone numbers identify recitals. Entries beginning with Article identify articles and their numbered paragraphs.}
	\label{fig:relevantParagraphSummary}
	\begin{tabular}{|l|p{2.7cm}|p{2.7cm}|}
		\hline
		\rowcolor{gray!20} 
		\textbf{Groups} & \textbf{Directed at Entities} & \textbf{Referencing Entities} \\
		\hline
		\textbf{Considerations} & 77; 83; 137; Article 23(3) ; Article 26(1); & 81; 82; 122; Article 20(1); Article 21(1); Article 21(3); Article 21(4); \\
		\hline
		\textbf{Scope} & 78; 79; & - \\
		\hline
		\textbf{Measures} & 58; 89; 96; 100; Article 21(2); & 49; 50; 51; 52; 80; Article 20(2); Article 24(1); Article 25(1); \\
		\hline
		\textbf{Supply Chains} & 85; & 90; 91; \\
		\hline
		\textbf{Third Parties} & 86; 88; & - \\
		\hline
		\textbf{Reporting} & 101; 102; 103; & 18; 105; 107; 120; Article 3(4); Article 23(1); Article 23(2); Article 23(4); Article 29(1); Article 29(2); Article 29(4); Article 30(1) \\
		\hline
	\end{tabular}
\end{table}

\subsection{Assessment of Focus Areas}
The map developed in the previous stage provides a unified reference of all NIS2 aspects relevant to the company’s context. This information is cross-analyzed with Celbi’s existing cybersecurity infrastructure to identify technical domains or organizational measures that require further development or reinforcement in alignment with NIS2. The assessment, conducted primarily by Celbi’s cybersecurity team, accounts for legal considerations, available resources, ongoing projects, and internal priorities to define and rank focus areas. This prioritization ensures an effective allocation of resources toward measures with the highest strategic and regulatory impact.

When considering all cybersecurity aspects in an IT infrastructure, the scope and reach of such a domain can be complex and overwhelming with many solutions to focus on. To support the prioritization process and communication with stakeholders, additional research on the modern threat landscape was conducted, highlighting phishing, ransomware, and supply chain attacks as some of the most relevant threats \cite{aag}. Reports of relevant authorities \cite{enisaThreatLandscape}\cite{enisaThreatLandscapeSupply} were also used to provide insights on potential future trends and security-related recommendations. 

\textit{Table \ref{fig:topicsToResearch}} presents the outcome of this stage. For each of the groups previously identified a series of detailed topics were identified, which were then researched by priority.

\begin{table*}[ht]
	\centering
	\caption{Identified focus areas and corresponding development priorities}
	\label{fig:topicsToResearch}
	\renewcommand{\arraystretch}{1.5}
	\begin{tabular}{|p{2.1cm}|l|p{11.9cm}|}
		\hline
		\rowcolor{gray!20} 
		\textbf{Groups} & \textbf{Priority} & \textbf{Artifact's topic/theme} \\
		\hline
		\textbf{Considerations} & Low & Mapping of NIS2 relevant term definitions, for transposition into internal documentation and procedures. \\
		\hline
		\textbf{Scope} & Medium & Review of the conducted risk assessments, to analyze whether they are current and if properly covered the scopes defined in NIS2. \\
		\hline
		\multirow{5}{*}[-0.8em]{\textbf{Measures}} & Medium & Zero Trust Principles. \\
		& Low    & Device Configurations. \\
		& High   & Secure Software Development, covering internal development and projects with third parties. \\
		& Medium & Human resources security and User access control policies. \\
		& Low    & Staff training and user awareness complemented by the profiling study (chapter 5) to help decision support. \\
		\hline
		\multirow{2}{2.1cm}[-0.5em]{\textbf{Supply Chains / Third Parties}} & Low     & Security of Communications, regarding encryption and used protocols. \\
		& Medium & Detailed mapping of NIS2 considerations and scope on this category to evaluate all internal scenarios in this context. \\
		\hline
		\textbf{Reporting} & Low  & Mapping requirements for internal documentation and procedures development considering NIS2 reporting obligations. \\
		\hline
	\end{tabular}
\end{table*}

\section{Case Study: Development Phase}
\label{sec:case_study_dev}
The final stage of the process focuses on developing the topics identified in the Assessment Phase. The \enquote{Development Phase} consists of generating a set of artifacts designed to support Celbi’s alignment with the \textit{NIS2 Directive}. Given the vastness of the cybersecurity domain, these artifacts—comprising documents of various types (e.g., graphs, charts, tables)—form a structured library of knowledge covering key technical and organizational areas. These topics are explored at different levels of detail, ranging from general overviews of a research topic to a detailed analysis of available solutions or functionalities provided in a specific ICT product. \textit{Table \ref{fig:artifactsIndex}} presents the produced artifacts, organized by the same groups previously identified and aligned with the identified topics. Compared to the artifacts planned in \textit{Table \ref{fig:topicsToResearch}} there were some small changes. The topic within the \textit{Scope} group was resolved during the research of higher priority topics. The topic of \textit{Security of Communications, regarding encryption and used protocols} was discarded since its area was already being covered by another Celbi's internal project. \textit{Human resources security and User access control policies} was eventually absorbed by the contents explored in the artifact developed in the context of topic \textit{Zero Trust Principles}. A total of 13 artifacts were developed in this work.

\begin{table*}[ht]
	\centering
	\renewcommand{\arraystretch}{1.5}
	\caption{Artifacts produced during the Development Phase. Numbers in the second column are artifact IDs used in the text. SDLC means software development life cycle. OWASP is the Open Worldwide Application Security Project. NIST is the National Institute of Standards and Technology. SSDF means Secure Software Development Framework. ICT and OT mean information and communication technology and operational technology, respectively.}
	\label{fig:artifactsIndex}
	\begin{tabular}{|p{2.5cm}|p{10cm}|}
		\hline
		\rowcolor{gray!20} 
		\textbf{Category} & \textbf{Artifact ID and Title} \\
		\hline
		\textbf{Considerations} & 1 -- NIS2 Definitions \\
		\hline
		\multirow{8}{*}{\textbf{Measures}} 
		& 2 -- Zero Trust Principle and Security Model \\
		& 3 -- Device Configuration \\ \cdashline{2-2}
		& 4.1 -- Security in Software Development Life Cycle \\
		& 4.2 -- Secure SDLC and the Organization \\
		& 4.3 -- Resources -- OWASP Mapping \\
		& 4.4 -- Resources -- NIST Secure Software Development Framework (SSDF) \\
		& 4.5 -- Resources -- Additional Secure Development Materials/Sources \\\cdashline{2-2}
		& 5 -- Awareness and Training \\
		\hline
		\multirow{3}{2cm}{\textbf{Supply Chains Third Parties}} 
		& 6.1 -- NIS2 considerations regarding Supply Chain and Third Parties \\
		& 6.2 -- ICT/OT supply chain's suppliers and providers Type \\
		& 6.3 -- Supply Chain Cybersecurity Good Practices \\
		\hline
		\textbf{Reporting} & 7 -- Reporting Obligations, Vulnerability Disclosure and Information Sharing \\
		\hline
	\end{tabular}
\end{table*}

This section will present each of the artifacts and discuss its relevance towards NIS2 compliance. The appendices contain the full versions of the artifacts. The main outcome of this process is a set of artifacts that provide the company with systematic information and practical guidance across key cybersecurity areas. Although they are tailored through the needs of Celbi, they can be equally relevant for a wide spectrum of entities, as they address several of the challenges inherent to cybersecurity. Although the level of detail in the artifacts could vary depending on a company's specific maturity and capabilities, the study contributes by identifying key challenges and harmonizes resources from various sources across various domains, generating publicly available artifacts offering actionable guidance for NIS2 compliance. For ease of reading, they will be presented in the order they are listed in \textit{Table \ref{fig:artifactsIndex}} and not in the order of development, which took the topics' priorities into consideration. 

\subsection{NIS2 Terms and definitions \#1}
NIS2 contains many terms and definitions used to designate specific services, key security terms, events, and other relevant aspects. During the initial analysis of NIS2, it was quickly identified by key stakeholders the need to somehow structure and codify these definitions. 

The process to create this artifact, which can be seen in \textit{Appendix \ref{app:terms_def}}, is divisible into two major steps: first all the relevant terms and definitions present in NIS2's article 6 “Definitions” are collected and structured (e.g., from simple concepts such as \textit{vulnerability} to \textit{national cybersecurity strategy}). The second step was to track all the external documents that were referenced in the set of definitions and include them in the document. This ensures that all the relevant information and full meaning of each term and definition is available in the artifact without the need for its reader to check additional documents. References in such cases were still included for source tracking if necessary.

Some of the terms snowballed beyond an externally referenced document, containing additional references to other documents. To keep the table easy to read and not overload it with potentially unnecessary details, these additional details were kept separated and can be consulted if needed. 
Keeping track of the full meaning of terms is crucial to avoid overlooking key aspects, especially when their scope is broad. This artifact ensures that a company's internal documentation and procedures comprehensively cover the full extent of a given term or service definition, aligning with its legal interpretation. By doing so, it strengthens the development of a compliant cybersecurity strategy.

\subsection{Zero Trust Principles \#2}

Zero Trust is a concept that has been evolving and growing in the last twenty years. Employing Zero Trust principles is one of the basic cyber-hygiene practices that entities should consider, as mentioned in NIS2's paragraph 89. The first step to develop artifacts related to this topic was to assess Celbi's awareness regarding the topic, to understand what areas would be relevant to focus on and target in the artifacts. This process was conducted, like in previous artifacts, through an interview with the cybersecurity team. Given the dimensions of the topic, the developed artifact should provide an overview of the evolution of the zero-trust notion, discuss related terms, while presenting available materials in terms of standards, tools and other relevant considerations. 

Important to note that the result of the contents for the artifact also covered and answered some of Celbi questions related to the topic \textit{Human resources security and User access control policies}. For this reason, this artifact, which can be seen in \textit{Appendix \ref{app:zta}}, is considered to satisfy both topics. Its contents are distributed across the following dedicated sections:

\begin{enumerate}
	\item \textbf{Zero Trust Principle and Security Model} – Insights on the term and concept of Zero Trust, as well as a brief look at its evolution history, to better understand this domain. 
	\item \textbf{Zero Trust Architecture (ZTA)} – Provide information on what constitutes a Zero Trust Architecture (ZTA), presenting the related NIST framework.
	\item \textbf{ZTA approaches} – Resumed overview of the different ways that an enterprise can enact a ZTA for workflows and different use cases.
	\item \textbf{ZT related terms, principles and tools.} – For awareness and training purposes, this section contains a glossary of terms related to Zero trust.
	\item \textbf{ZTNA market overview and guidelines} – Presents additional info gathered during the artifact development related to the available market of ZTNA products and relevant guidelines important to consider when assessing such products.
	\item \textbf{Identity and Access Management (IAM)} – Resumed overview of IAM, offering a starting point on developing a IAM strategy, 
	\item \textbf{Microsoft Entra Solutions for IAM} – Provide an overview of the resources and tools currently available in Microsoft Entra in the context of IAM. 
\end{enumerate}

This artifact provides guidance on the various terms and solutions related to Zero Trust principles, outlining key aspects to consider when implementing them. By understanding the history and evolution of the term, recommended architectures, and available solutions, organizations can better evaluate their specific requirements and determine the most suitable approach to adopt.

\subsection{Devices Configuration \#3}

Acquired devices, products and systems are usually complex when considering the number of functionalities and configurations that they allow. Such options can be left on default settings due to the lack of time or knowledge to properly explore the product, which can lead to vulnerabilities due to misconfiguration or forgetting to change the default settings. This artifact, \textit{Appendix \ref{app:device_config}}, presents the CIS Benchmarks as a proposed solution for this context while demonstrating how the CIS Benchmarks platform and respective documents are structured. Even smaller companies rely on diverse IT equipment, which can introduce security gaps when default settings remain unchanged or configurations are incomplete. These benchmarks, tailored to the specifications of each device and technology, help mitigate such risks while streamlining the company's onboarding procedures for new equipment. This approach reduces the need for manual exploration of all configuration options and functionalities, ensuring a more secure and efficient deployment.  

\subsection{Secure Software Development \#4.1 to \#4.5 }

Entities should consider many aspects of their information systems security in terms of acquisition, maintenance and also in their development (Article 21(2)(e) NIS 2 Directive). The group's IT infrastructure contains application teams responsible for developing projects throughout the year in order to develop application solutions that can be consumed internally or in the context of partnerships with external suppliers and third parties. 

This artifact covers the topic of \textit{Secure Software Development} described in \textit{Table \ref{fig:topicsToResearch}}. However, this domain is very broad and comprises several stages, each with its level of technical and practical considerations. Hence, the first step was understanding the awareness level of Celbi's cybersecurity team and key users on this topic. Also of key importance was to understand the development environment and the methodologies and processes currently used by Celbi's development teams. Internal interviews were conducted with key users and the gathered insights were later discussed with the internal cybersecurity team for alignment on which subtopics this artifact should focus on. 

The first step involved identifying which guidelines and initiatives were being developed within the EU in this field. Earlier ENISA studies \cite{enisaSecureSoftwareInits} noted a lack of government-driven initiatives in secure software engineering across the EU, in contrast with the US. More recent reports \cite{enisa_softReport2020} have highlighted overlapping standards, certification challenges, and the difficulty non-technical entities face in verifying software security. The implementing act for the Common Criteria-based EU cybersecurity certification scheme (EUCC) \cite{regulation2019881} partially addresses these issues by establishing the first EU-wide certification scheme \cite{UEcertificationScheme}.

\begin{enumerate}
	\item \textbf{4.1 - Security in Software Development Life Cycle} – For the goal of awareness, this artifact, which can be seen in \textit{Appendix \ref{app:sdlc}}, presents an overview of the process of software development and how it can be complemented with cybersecurity.
	\item \textbf{4.2 - Secure SDLC and the Organization} – Building on 4.1, this artifact, available in \textit{Appendix \ref{app:sdlc_orgn}} by presenting the role and tasks from an organization perspective to integrate and maintain secure practices across the many iterations of cycle development.
	\item \textbf{4.3 - Resources – OWASP Mapping} – Overview and map of the works and resources from OWASP community that can help with secure software development practices, \textit{Appendix \ref{app:sldc_mapping}}.
	\item \textbf{4.4 - Resources – NIST Secure Software Development Framework (SSDF)} – In this artifact, \textit{Appendix \ref{app:sldc_nist}}, the content of NIST SSDF, while correlating it with the materials present on 4.1 and 4.2 a establish a map of resources and sources on which NIST SSDF is based on, providing in this way, additional frameworks and documents in the context of Secure Software Development.    
	\item \textbf{4.5 - Resources – Additional Secure Development Materials/Sources} – Additional sources for secure software development \textit{Appendix \ref{app:sldc_additional}}.
\end{enumerate}

Secure software development was the only focus area rated high priority in Celbi's assessment and covered both internal development and projects with third parties (Table~\ref{fig:topicsToResearch}). To address the development aspect of Article~21(2)(e) of the NIS2 Directive, artifact~4.4 (Appendix~\ref{app:sldc_nist}) maps NIST SSDF tasks to the lifecycle stages described in artifacts~4.1 and~4.2. For example, the study maps task~PW.5.1, which calls for secure coding practices suited to the development languages and environment, to stage~A3 in artifact~4.1, where the team carries out the planned development work.

By providing a broad overview of all relevant aspects of SSE, rather than focusing on a specific software development methodology, this artifact remains universally applicable. Additionally, it compiles valuable sources of information, serving as both an introduction to the topic and a resource for integrating security as a fundamental aspect throughout all stages of the software development process.

\subsection{Awareness and Training \#5}

The \textit{NIS2 Directive} includes many considerations that cover the need for different types and levels of awareness, along with specific measures and considerations regarding training. Many of them, although not necessarily representing an obligation or a measure to implement by an essential or important entity, clarify how certain processes can contribute to awareness. This can further help reveal additional sources of information to promote an internal training and awareness development strategy.

This artifact, which can be seen in \textit{Appendix \ref{app:awareness}}, starts by presenting all the NIS2 relevant considerations toward training and awareness. These details are organized in a table referencing the relevant sections of the \textit{NIS2 Directive}, identifying to which entity or stakeholder is the content directed to and its description. Additional insights concerning the processing of personal data in the context of enhancing awareness and training activities are also included in the table. Relevant keywords are also included and highlighted to help guide the reading. 

In addition to an overview of training and awareness considerations in NIS2, the artifact also contains a list of valuable documents that should be consulted to further guide Celbi's cybersecurity strategy. These are a set of reports and bulletins that are published with certain recurrence and provide valuable insights on matters such as the national and international threat landscape, publication of relevant new regulations, detailed reports on specific topics of growing relevance and others.

These materials are not the only source with potential value. Both CNCS and ENISA entities have other dedicated reports, covering areas and topics that may bring additional value. Some of those publications are specific to events or themes and may have only a single version/publication, like a specific assessment regarding the impact of a new regulation, or a specific new vector/trend of threat recently monitored. 
Along with the already covered information available in the artifact, it also includes a recommendation on how this information can be used by Celbi:
an internal periodic procedure should be established, to keep track of relevant publications. The study of such sources may provide additional guidance on what areas and contents to focus on for the training sessions. This can be tailored to specific user roles or company site locations in accordance with the nature of the reported threats and rising trends. Furthermore, these publications can additionally provide guidance and support decisions on other areas of Celbi's cybersecurity strategy, such as new policies to watch for, measures to deploy or specific areas or technologies to focus further assessments/developments. The periodicity of the mentioned procedure should be defined considering the publishing recurrence of the relevant materials. 

Through this artifact, companies operating in the EU will gain access to additional resources that support decision-making when determining which cybersecurity areas or domains to prioritize for future investment. It also provides a summary of NIS2 considerations related to awareness and training. While some identified materials are specific to Celbi's market, companies can apply the same approach by consulting their respective national authorities for additional bulletins and resources to further enhance their cybersecurity strategies.

\subsection{Supply Chain and Third Parties}

Supply chain and third parties are becoming one of the major attack vectors \cite{aag}. NIS2 contains several considerations focusing on risk management from supply chain and third parties/service providers (recitals 85, 86, 88, 90, 91). Considering the nature of the various daily activities of the group's business process, these recitals raised a set of questions and possible implications that might be currently overlooked. The first step was to thoroughly review the directive for every article and paragraph concerning supply chain and third parties, ensuring the identification of all valuable insights related to this matter. Two artifacts were produced as a result of the analysis conducted in this topic: 

\begin{enumerate}
	\item \textbf{6.1 – NIS2 considerations regarding Supply Chain and Third Parties} – This artifact, \textit{Appendix \ref{app:supply_chain}}, aims to map NIS2 considerations is the result of the search conducted.
	\item \textbf{Artifacts 6.2 and 6.3 – ICT/OT supply chain's suppliers and Supply Chain Cybersecurity Good Practices} – Building on 6.1, both these artifacts, \textit{Appendix \ref{app:supply_chain_types}} and \textit{Appendix \ref{app:supply_chain_practices}},  investigate what EU standards and/or solutions were available to help handle risk management toward Supply Chain and Third Parties.
\end{enumerate}

This artifact provides a foundational template, aligned with NIS2 recommendations, to help organizations develop a cybersecurity strategy tailored to this often-overlooked aspect

\subsection{Reporting Obligations and Disclosure \#7}

The initial plan for this topic was an artifact that would simply present the obligations of entities such as Celbi, concerning reporting and disclosure. For this purpose, a search using the keywords \enquote{report} and \enquote{disclosure} was conducted on the directive in order to identify all relevant recitals. This process, however, revealed additional points of interest such as specification on what will be the processes through which information will be handled by the competent authorities, or what destiny is given to the flow of information during the disclosure of an incident or vulnerability. As a result, the scope of this artifact was broadened.

The resulting artifact, \textit{Appendix \ref{app:reporting_obligations}}, is divided in two parts. The first part seeks to respond to several questions, such as what will be expected or encouraged to be reported by entities, which potential obligations will be set in this context and what structures and processes will accommodate these procedures. The contents are then distributed across 3 major groups:
\begin{itemize}
	\item \textbf{Cybersecurity Measures and Vulnerabilities disclosure} – With headers color-coded green, this group presents the considerations regarding information and vulnerability disclosure.
	\item \textbf{Incident Reporting} – Color-coded orange headers, this group lists the obligations related to the procedure needed to take during the process of reporting an incident.
	\item \textbf{Voluntary Notification and Information Sharing} – With headers color-coded blue, this group lists considerations regarding the voluntary sharing of information. 
\end{itemize}

For each content its original location in NIS2 is also identified and, for each row, two columns separate considerations directed at entities from those directed at Member States. The reasoning behind this chosen presentation is that some of the NIS2 recitals and points include some remarks relevant and directed to more than one type of stakeholder, which raises some challenges when it comes to presenting these materials in an organized and intuitive way. 

The second part of the artifact presents the remaining considerations identified during the initial search that may affect the disclosure of information among different stakeholders or are in some way related to it. Other contexts such as guidelines for data processing between authorities and Member States, stakeholder obligations regarding vulnerability disclosures, legal implications for actors disclosing information, among others, are also presented in this second part. The resulting table contains a total of 23 items and like previous artifacts, it describes where the content can be found in NIS2 and also the type of stakeholder to which the item is directed. 

Through this artifact, entities will find a comprehensive summary of NIS2's considerations regarding information sharing, organized into categories such as vulnerability disclosures, incident reporting, voluntary notifications, and information sharing. It also clarifies the roles and responsibilities of the relevant stakeholders, offering deeper insights into their respective duties and contributions to the cybersecurity ecosystem.

\section{Discussion}
\label{sec:discussion}
Despite the growing importance of cybersecurity legislation, interpreting and operationalizing legislative frameworks remains a significant challenge. NIS2 represents one of the most recent and far-reaching European cybersecurity directives, with a broader scope and deeper implications than its predecessors. It is an extensive and complex legal instrument whose requirements reference numerous complementary documents. For instance, a seemingly straightforward topic such as secure software development involves diverse resources from trusted sources including NIST, OWASP, and ENISA. Understanding how these sources interrelate, what assets they provide, and which are most applicable is far from trivial.

This paper presented a detailed analysis of NIS2 using a structured multi-stage methodology to assess its impact within the context of a large European company, Celbi. The work addressed the challenges of interpreting such a comprehensive directive and proposed a systematic approach to understanding its implications for entities falling under NIS2’s scope. At Celbi, the method helped identify priorities for cybersecurity planning while national transposition was pending in Portugal.

The Celbi case documents how the method translates provisions of the directive into guidance for technical and organizational decisions. The mapping in Table~\ref{fig:relevantParagraphSummary} summarizes 29 recitals and 18 article references across six categories. Assessing the mapped provisions against Celbi's existing practices and priorities yielded 10 focus areas (Table~\ref{fig:topicsToResearch}). Secure software development was the only area rated high priority. The development phase produced 13 artifacts (Table~\ref{fig:artifactsIndex}), including guidance on secure development, supplier risk, staff awareness, zero trust, and incident reporting.

The case study strengthened Celbi's basis for cybersecurity planning by linking relevant NIS2 provisions to priorities reviewed with its IT team and to guidance for addressing them. For example, the supply-chain artifacts identify types of ICT and OT suppliers and compile practices Celbi can use when reviewing supplier risk. The reporting artifact links incident reporting and vulnerability disclosure provisions to their sources in the directive, giving staff a reference when reviewing internal procedures.

In a case study of two Belgian medical-supply manufacturers, de Wit~\cite{dewit2025nis2} used the CyberFundamentals self-assessment to rate how security measures were implemented and documented. Both companies scored higher on implementation than on documentation. This illustrates how a self-assessment can identify gaps in existing practices, while the Celbi case illustrates how mapping the directive's provisions can help prioritize topics and develop guidance for addressing them.

The study did not collect before-and-after data on cybersecurity maturity, audit findings, or incident frequency. The effect of implementing the guidance on those outcomes therefore remains unmeasured.

Celbi's legal department reviewed the initial scope assessment. The resulting mapping and artifacts provide technical and organizational guidance based on the EU directive.

\section{Threats to Validity}
\label{sec:threats}

Although this work followed a structured and carefully planned methodology, several threats to validity must be acknowledged.

The method was applied at a single industrial site within a large enterprise. The assessment drew on access to Celbi's internal processes, input from IT staff when reviewing current practices and priorities, and input from legal staff when reviewing how the directive might apply to Celbi. These conditions shaped the depth of the assessment and the priorities and artifacts it produced. Results may vary with an organization's size, cybersecurity maturity, and available expertise.

Smaller organizations could reuse the steps for mapping relevant NIS2 provisions, reviewing current practices, and prioritizing focus areas. The depth of that review will depend on access to internal documentation and staff who know the organization's systems and processes. Reviewing how the directive applies to the organization may also require access to legal expertise.

The case study used the EU directive as its reference. The findings therefore concern the technical and organizational implications of the EU text. Assessing Celbi's obligations or the legal sufficiency of specific measures under Portugal's subsequent transposition would require a separate assessment.

\section{Conclusion}
\label{sec:conclusion}
Modern economies depend on highly interconnected ICT systems that support critical infrastructures and high-value markets, making them increasingly attractive targets for malicious actors. To address the growing frequency and impact of cyberattacks, the European Union has introduced legislative instruments aimed at improving cybersecurity and resilience. Among these, the \textit{NIS2 Directive} establishes a comprehensive framework to strengthen risk management and governance practices across sectors. However, its legal language and conceptual scope can be difficult to interpret from a technical or operational standpoint.

This work analyzed NIS2 from a technical perspective, demonstrating how such a legislative instrument can serve as a mechanism for structuring and enhancing organizational cybersecurity strategies. Through a systematic methodology applied to a large European enterprise, the study explored how NIS2 principles can be translated into concrete technical and organizational actions, rather than treated purely as compliance obligations. The process resulted in a set of artifacts and methodological steps that consolidate guidance from multiple authoritative sources and map them to the directive’s high-level objectives.

The proposed approach emphasizes understanding, alignment, and operationalization—helping organizations anticipate future regulatory expectations while proactively improving their cybersecurity posture. The resulting artifacts, made publicly available, provide reusable resources that illustrate how policy-driven frameworks such as NIS2 can be leveraged to guide structured, evidence-based cybersecurity development across diverse contexts.

The study relates to two United Nations Sustainable Development Goals~\cite{un2030agenda}. Its guidance on managing cybersecurity risks in industrial operations addresses infrastructure resilience (Goal~9), and its guidance on reporting incidents to competent authorities supports accountability (Goal~16).

Applying the method in organizations of different sizes and sectors would help identify which steps and artifacts can be reused across settings. Future studies could also examine how the method fits into existing cybersecurity management systems and compare how AI tools and human auditors map the same evidence to applicable requirements.

\section{Open Science Expectations}

Although this study was developed within an industrial context in collaboration with a university, all artifacts/resources were developed considering its public disclosure. All the base documents and materials used in the study are also publicly available and referenced through the produced artifacts which will be made publicly available.

\appendices

\includepdf[pages={1}, pagecommand={\section{NIS2 Terms and Definitions}\label{app:terms_def}}]{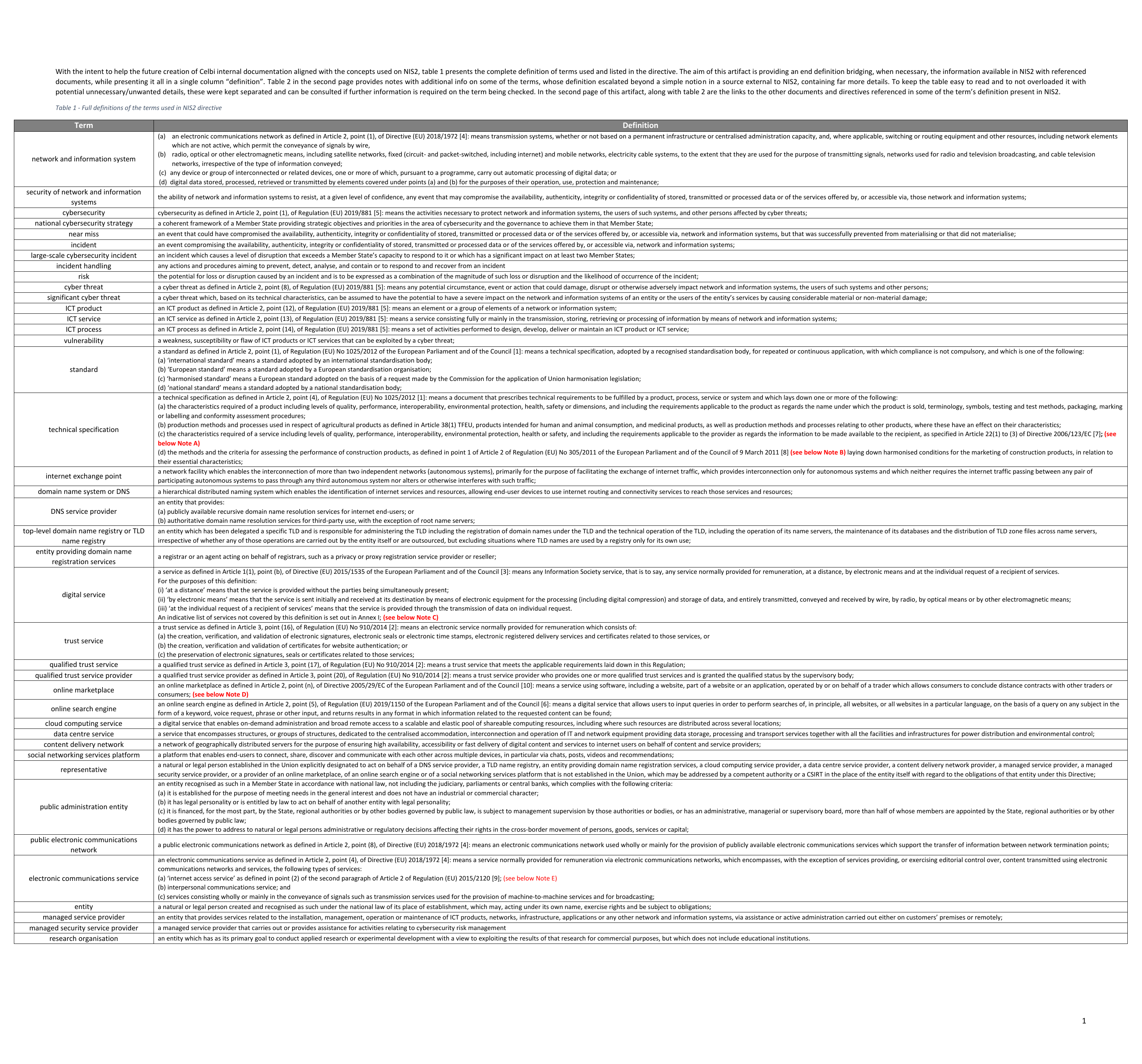}
\includepdf[pages={2}]{PDFs/1_NIS2_Definitions.pdf}

\includepdf[pages={1}, scale=0.9, pagecommand={\section{Zero Trust Principle and Security Model}\label{app:zta}}]{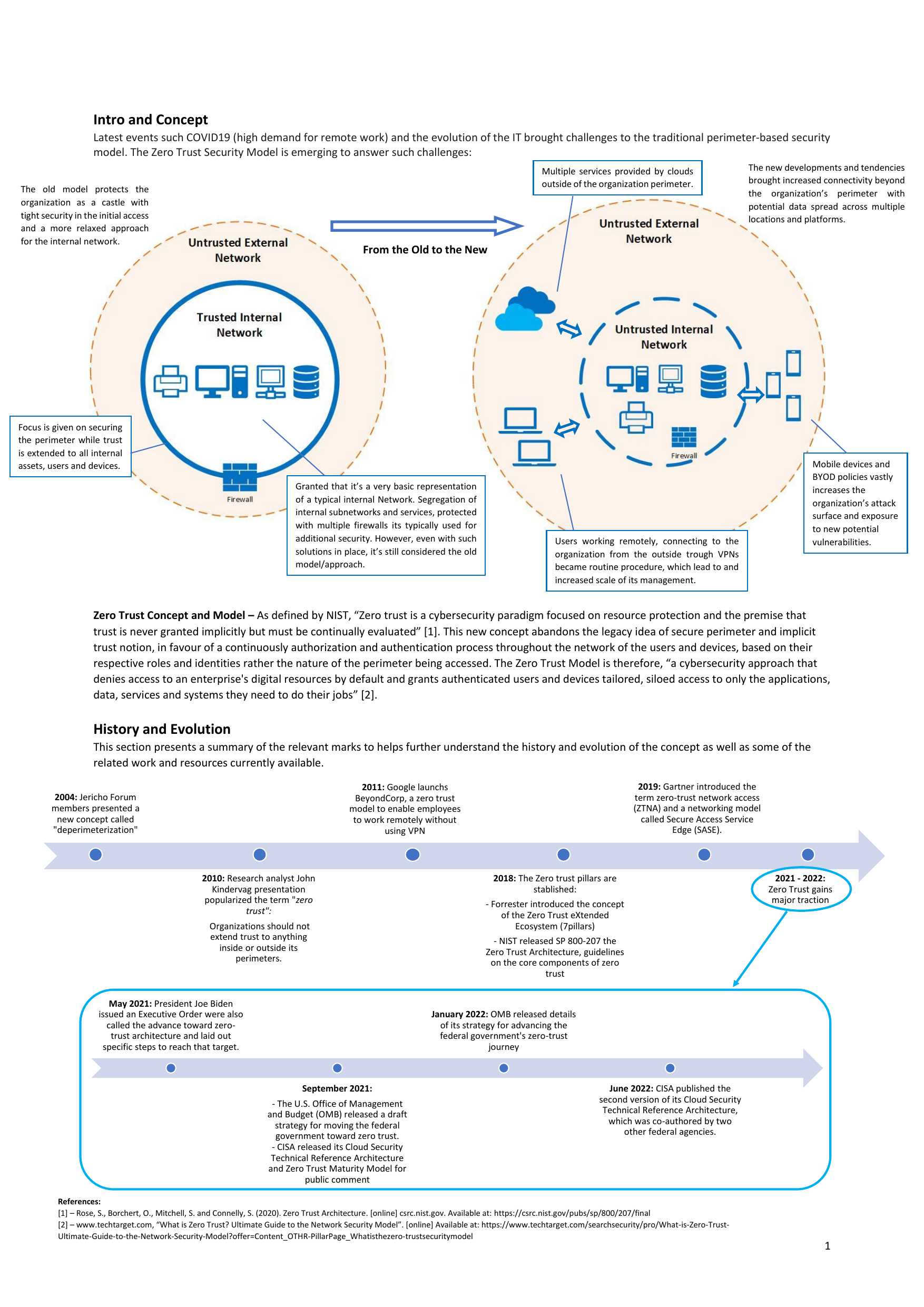}
\includepdf[pages={2-8}]{PDFs/2_Zero_Trust_Principle_and_Security_Model.pdf}

\includepdf[pages={1}, scale=0.9, pagecommand={\section{Device Configuration}\label{app:device_config}}]{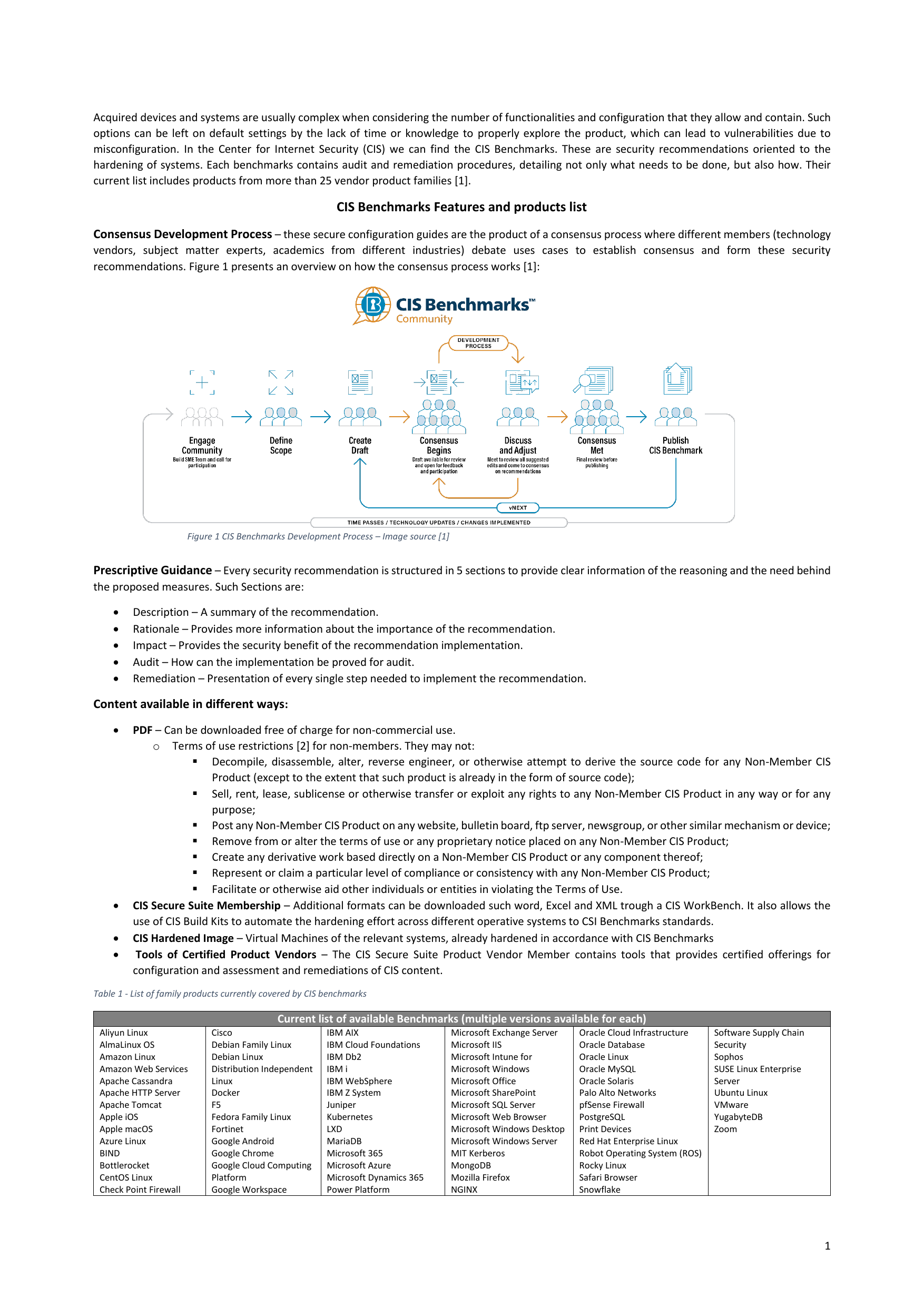}
\includepdf[pages={2}]{PDFs/3_DeviceConfiguration.pdf}

\includepdf[pages={1}, scale=0.9, pagecommand={\section{Security in Software Development Life Cycle}\label{app:sdlc}}]{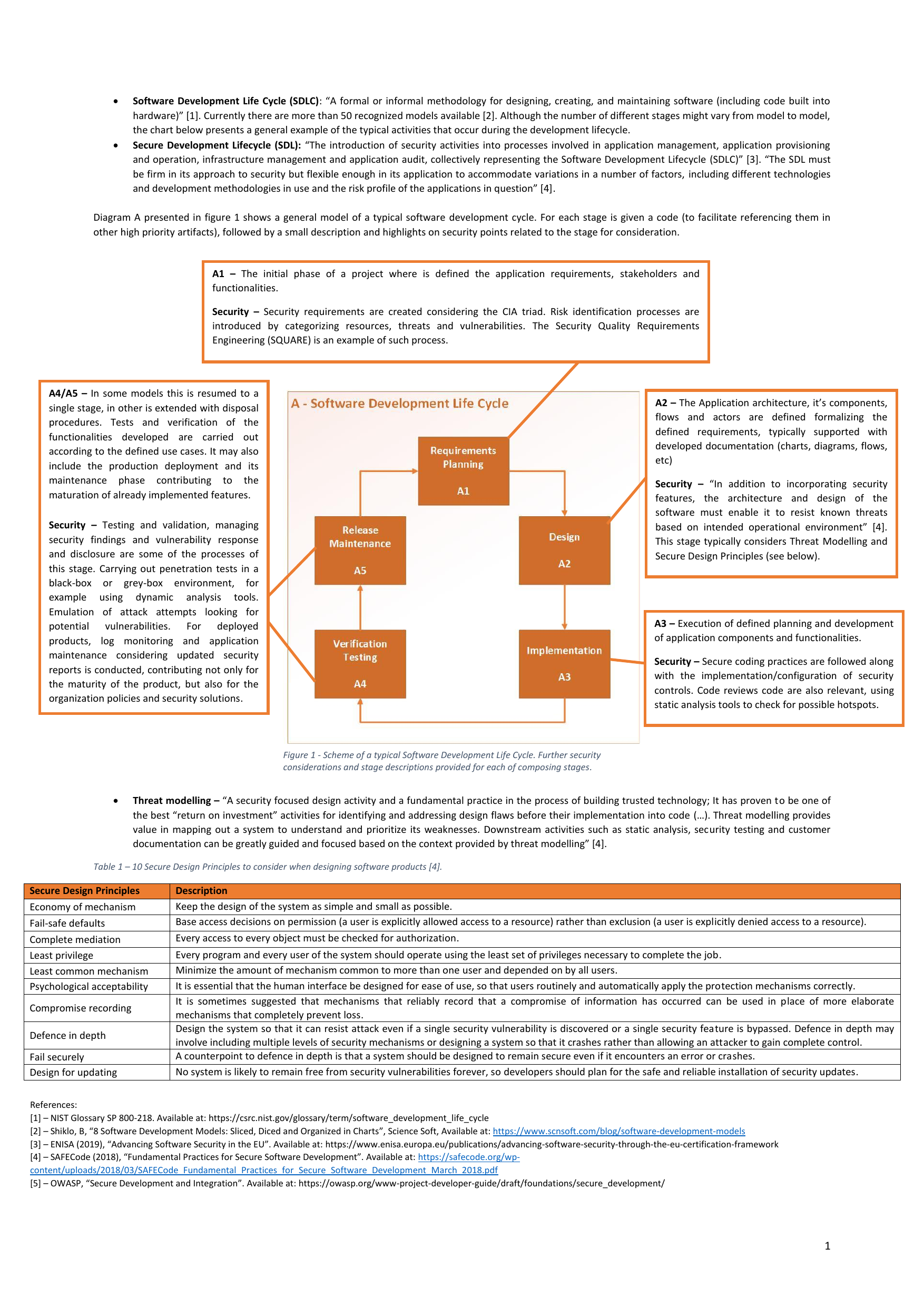}

\includepdf[pages={1}, scale=0.9, pagecommand={\subsection{Secure SDLC and the Organization}\label{app:sdlc_orgn}}]{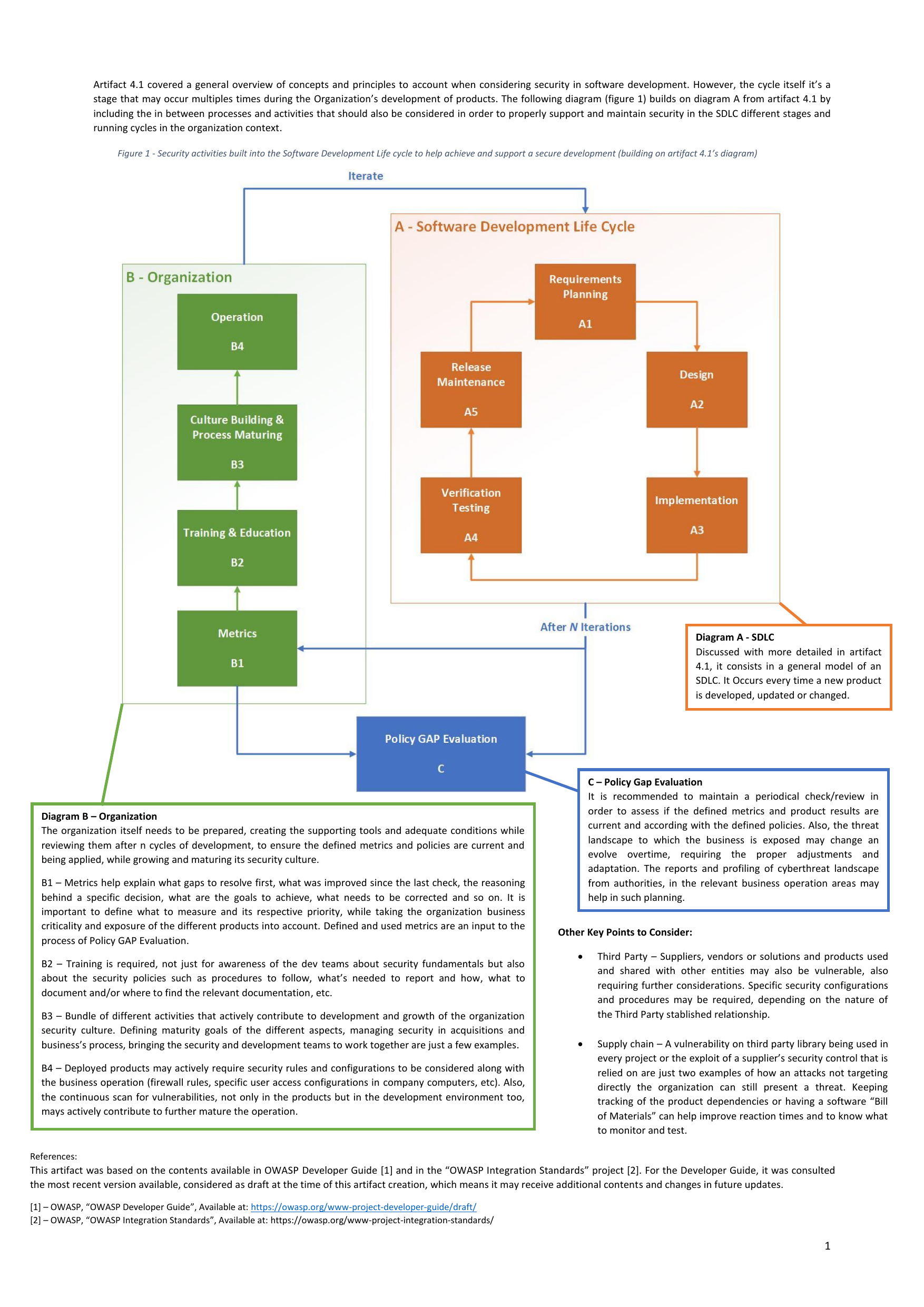}

\includepdf[pages={1}, scale=0.9, pagecommand={\subsection{OWASP Mapping}\label{app:sldc_mapping}}]{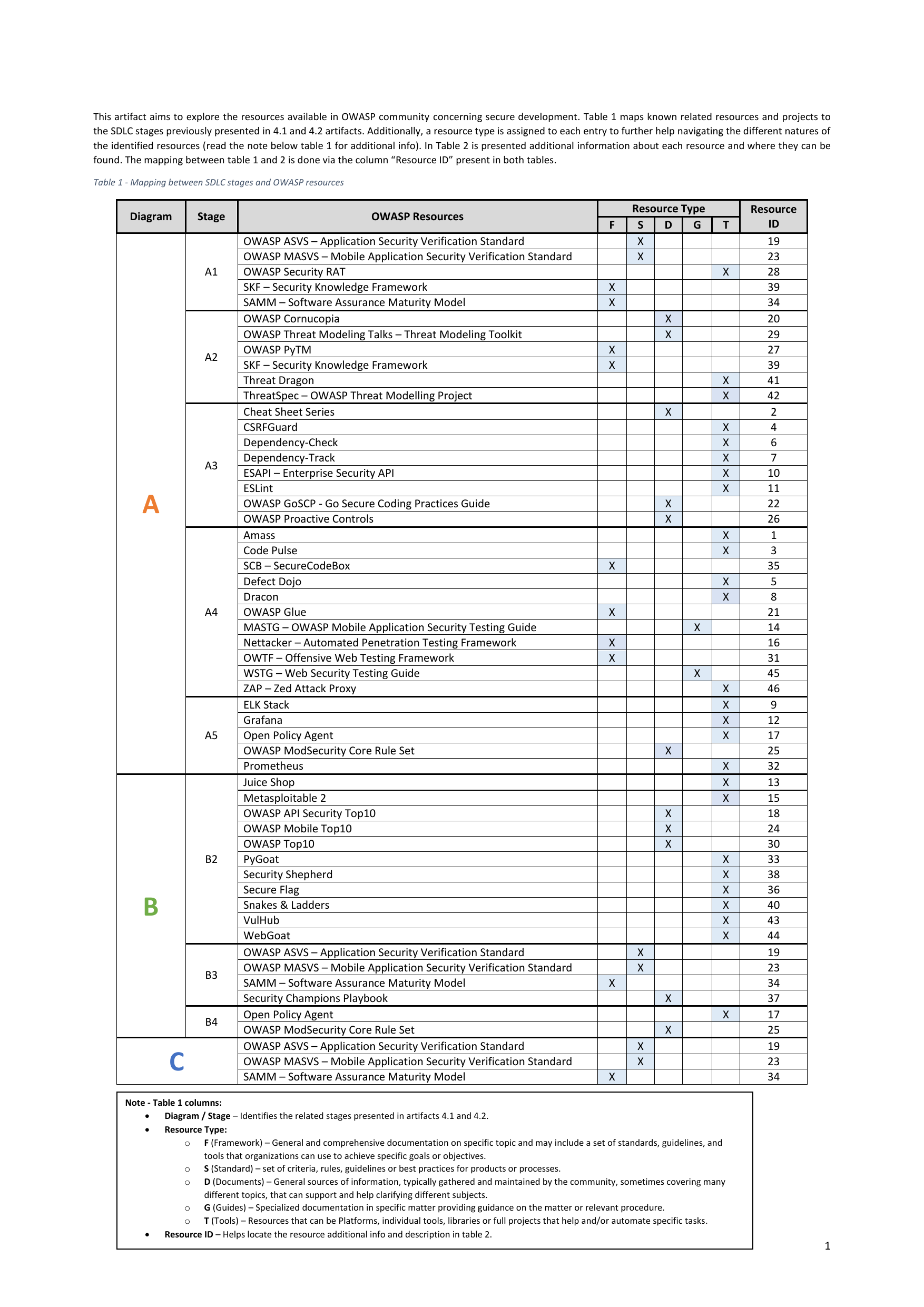}
\includepdf[pages={2-3}]{PDFs/4.3_OWASP.pdf}

\includepdf[pages={1}, scale=0.9, pagecommand={\subsection{NIST Secure Software Development Framework (SSDF)}\label{app:sldc_nist}}]{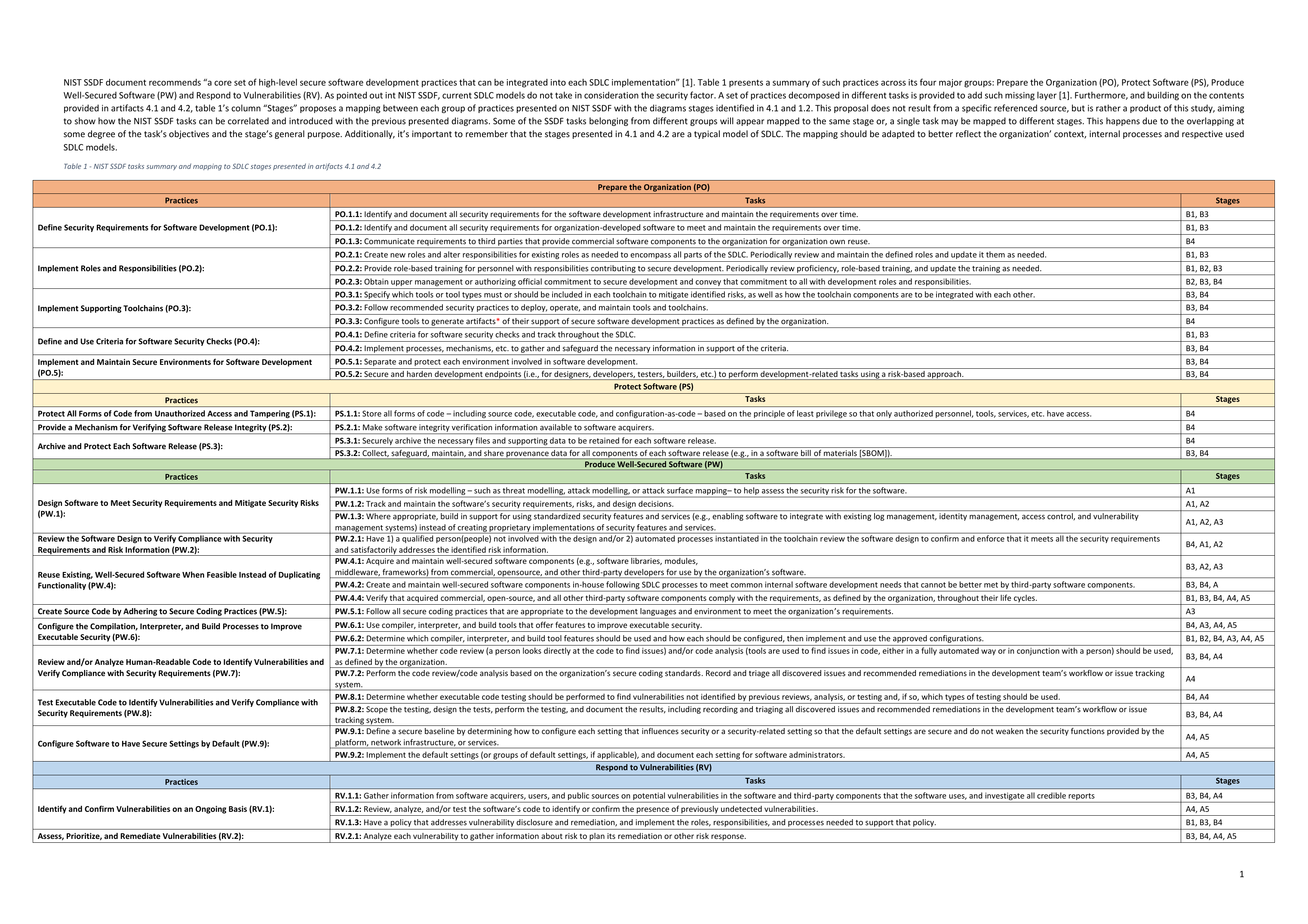}
\includepdf[pages={2-3}]{PDFs/4.4_NIST.pdf}

\includepdf[pages={1}, scale=0.9, pagecommand={\subsection{Additional Secure Development Materials/Sources of Information}\label{app:sldc_additional}}]{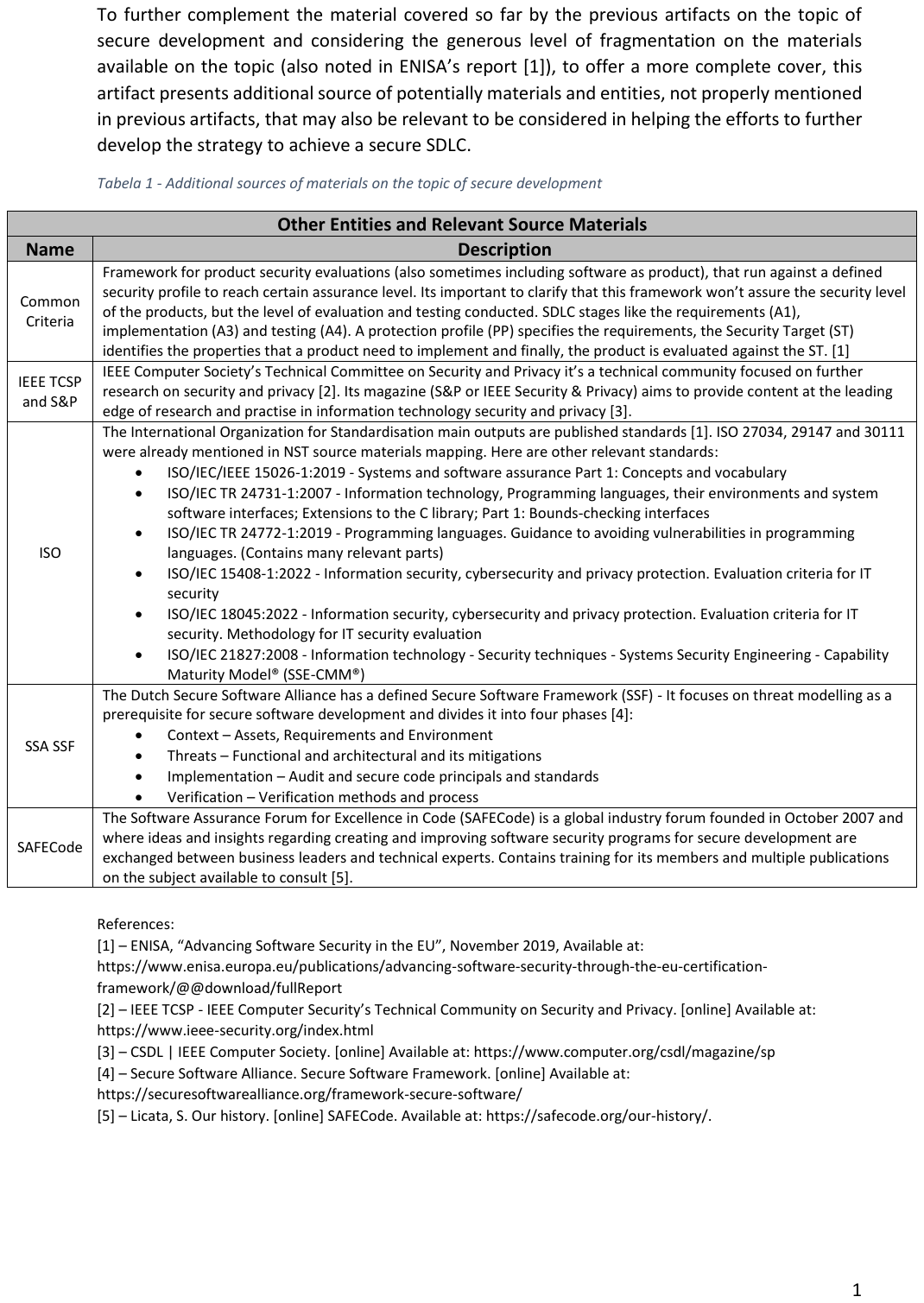}

\includepdf[pages={1}, scale=0.9, pagecommand={\section{Awareness and Training}\label{app:awareness}}]{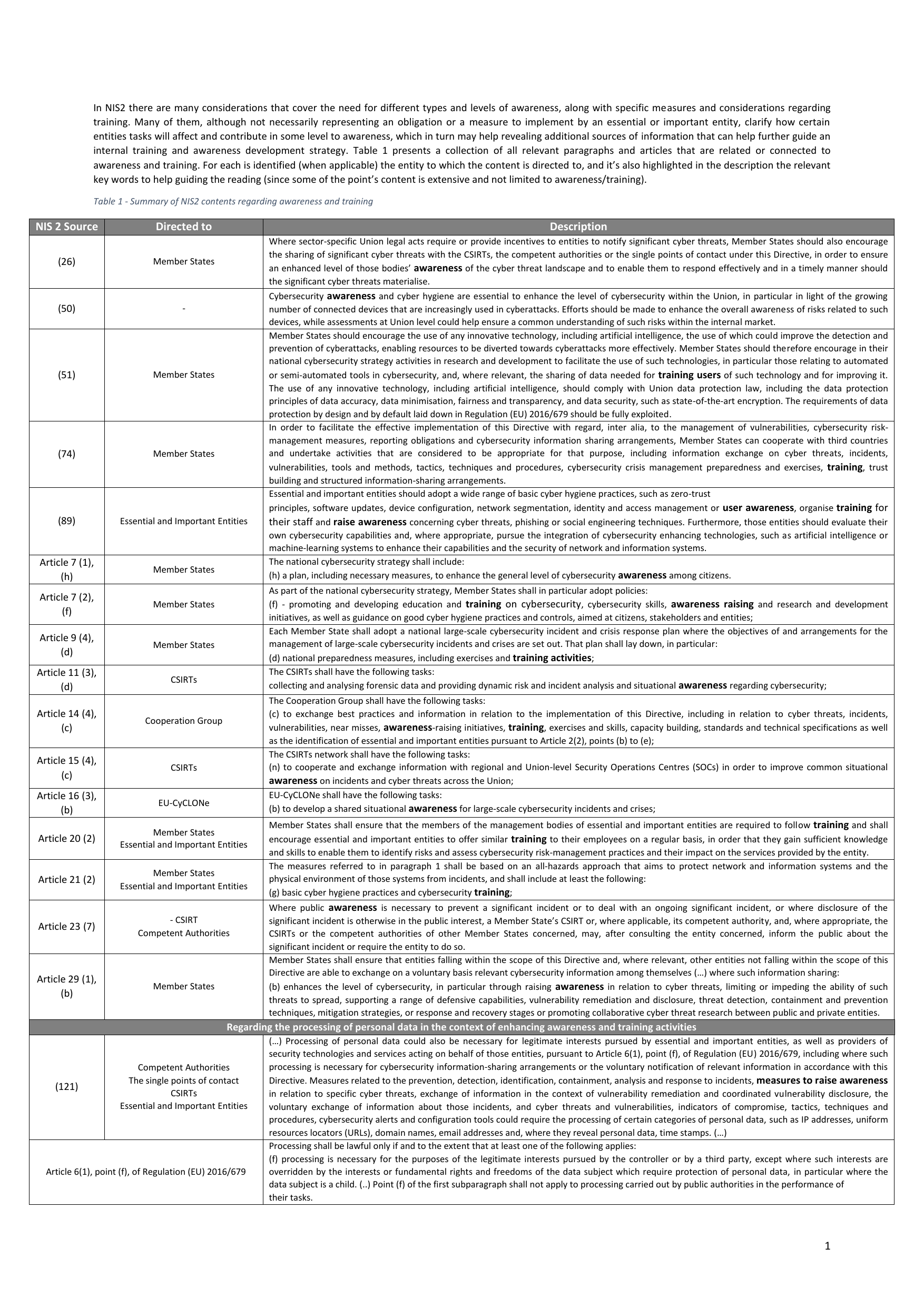}
\includepdf[pages={2}]{PDFs/5_Awareness_Training.pdf}

\includepdf[pages={1}, pagecommand={\section{NIS2 considerations regarding Supply Chain and Third Parties}\label{app:supply_chain}}]{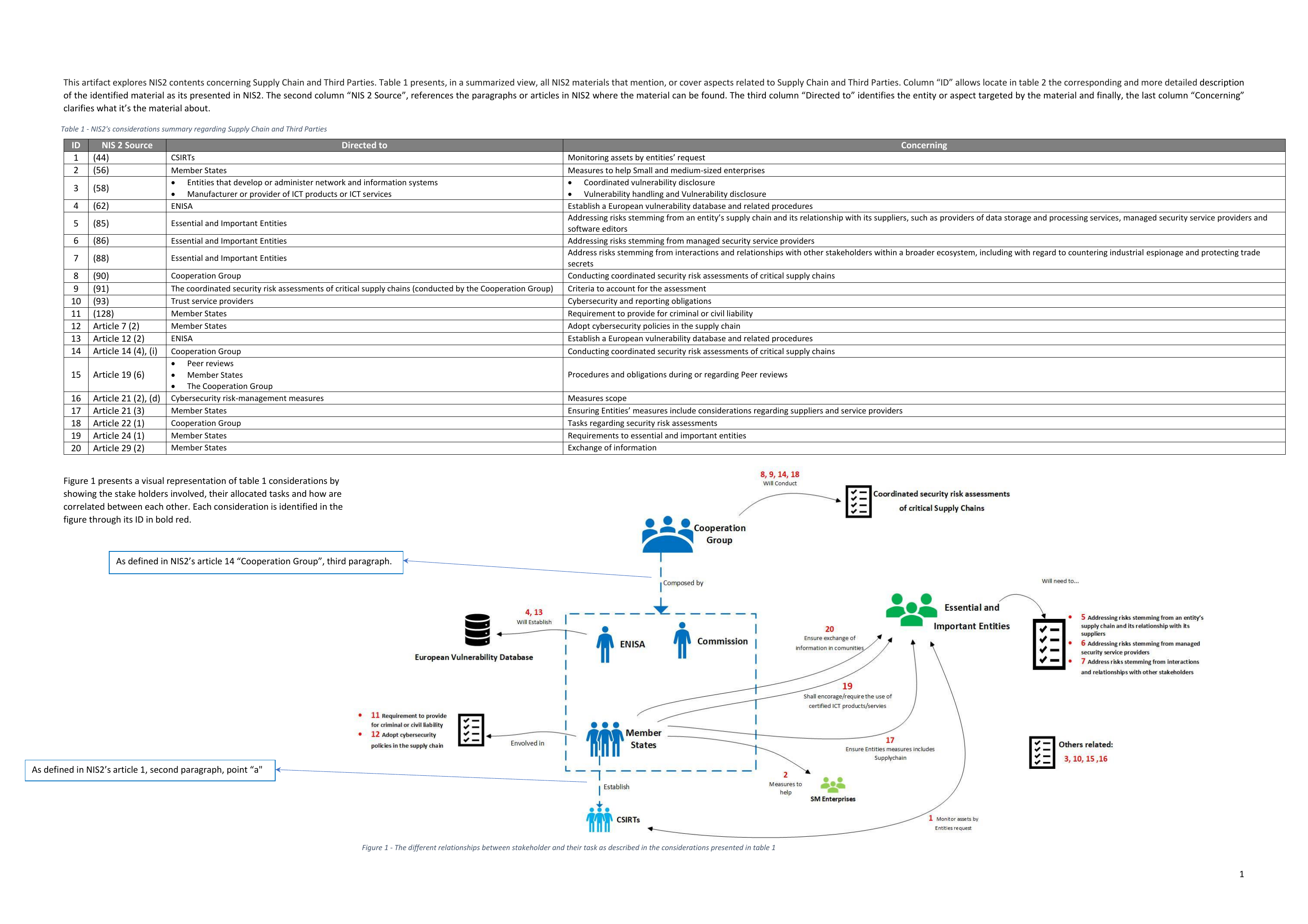}
\includepdf[pages={2}]{PDFs/6.1.pdf}

\includepdf[pages={1}, scale=0.9, pagecommand={\subsection{ICT/OT supply chain's suppliers and providers Type}\label{app:supply_chain_types}}]{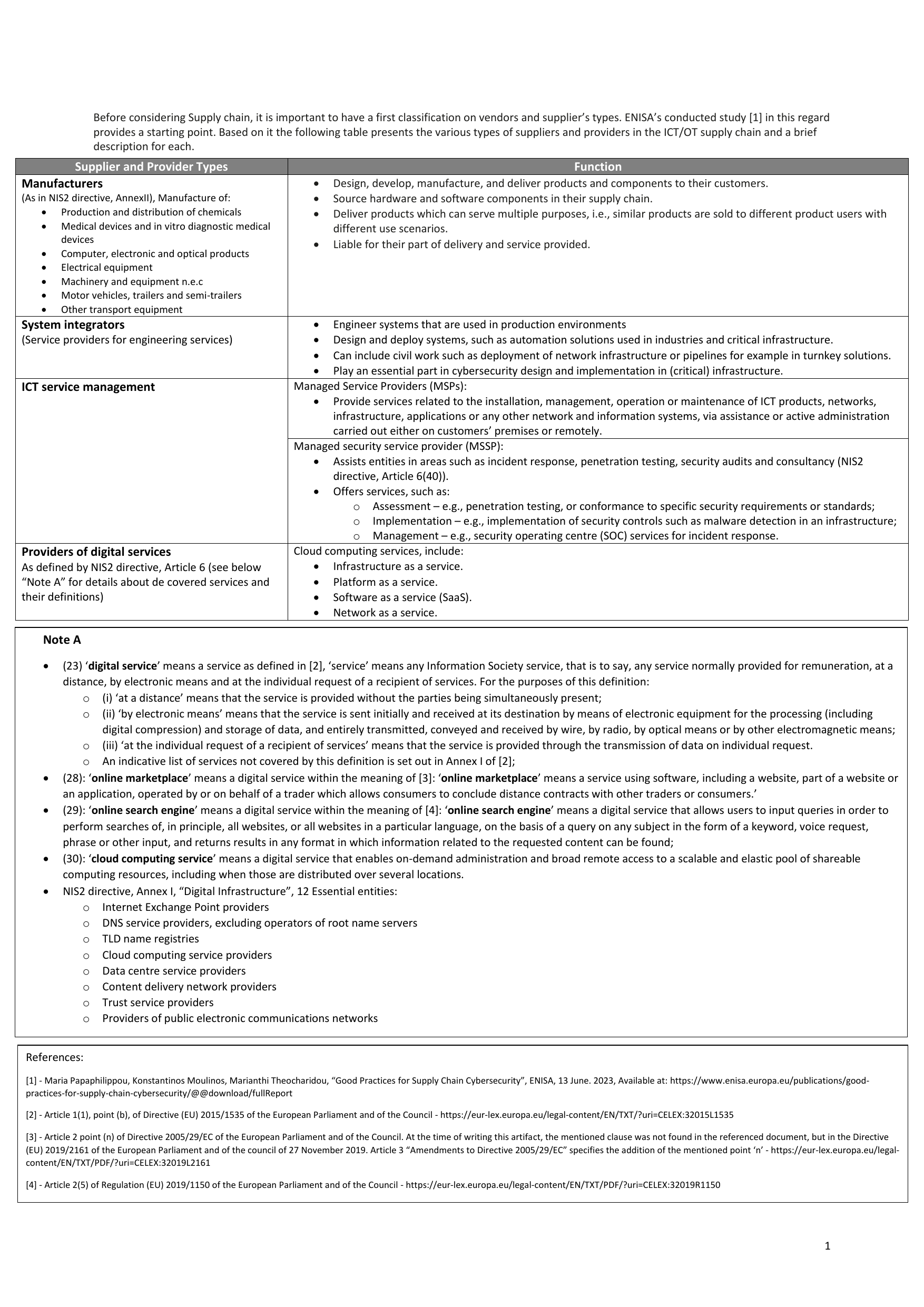}

\includepdf[pages={2}, pagecommand={\subsection{Supply Chain Cybersecurity Good Practices}\label{app:supply_chain_practices}}]{PDFs/6.2_6.3.pdf}
\includepdf[pages={3-5}]{PDFs/6.2_6.3.pdf}

\includepdf[pages={1}, pagecommand={\section{Reporting Obligations, Vulnerability Disclosure and Information Sharing}\label{app:reporting_obligations}}]{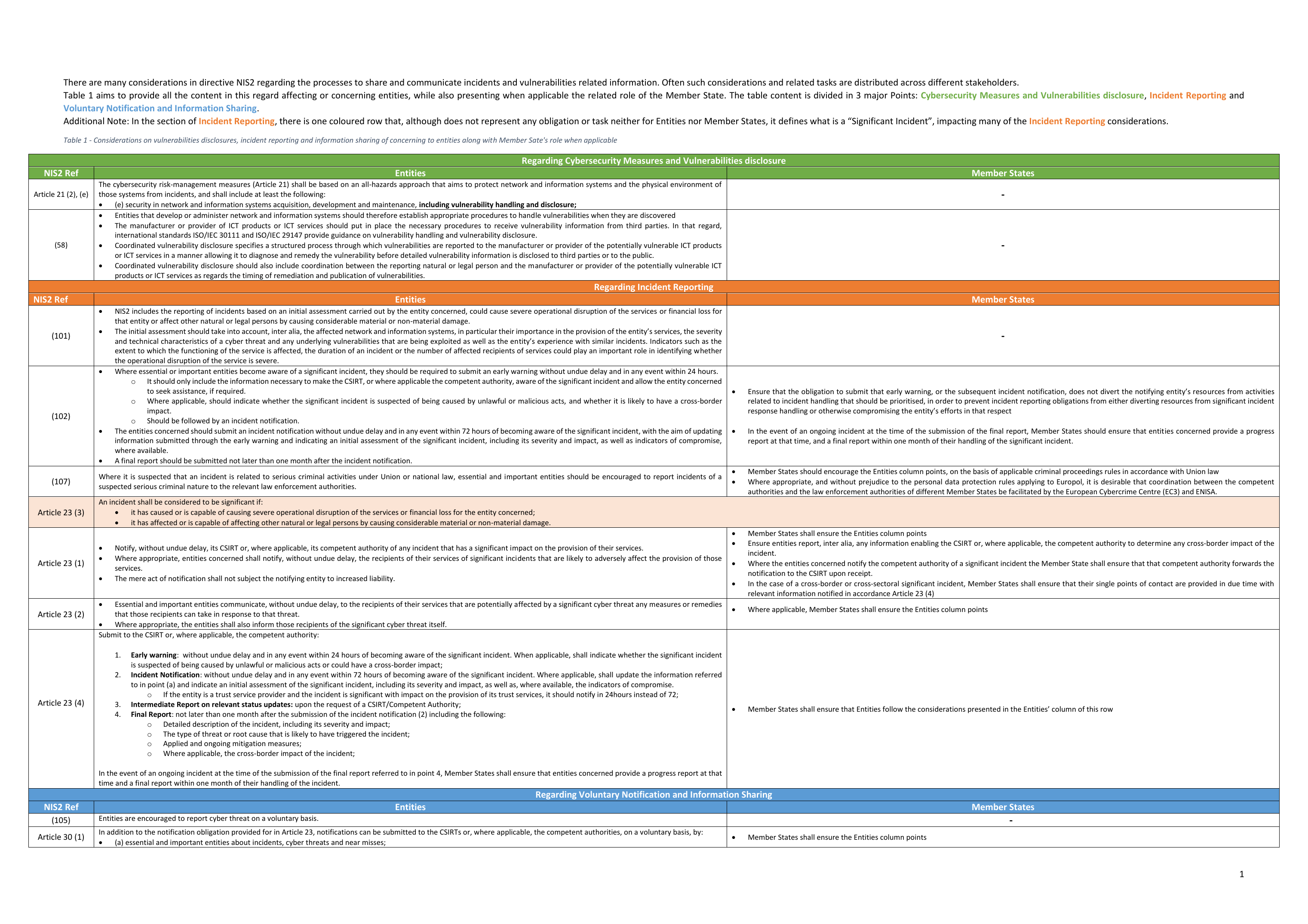}
\includepdf[pages={2-3}]{PDFs/7_ReportObligations_VulnerabilitiesDisclosure.pdf}

\bibliographystyle{IEEEtran}
\bibliography{IEEEabrv, references}

\end{document}